\documentclass[a4paper,aps,prd,twocolumn,groupedaddress]{revtex4-2}
\usepackage{natbib}
\usepackage{dcolumn}
\usepackage{mathtools}
\usepackage{amsmath}
\usepackage{amssymb}
\usepackage{amsfonts}
\usepackage{mathrsfs}
\usepackage[caption=false]{subfig}
\usepackage{graphicx}
\usepackage{bm}
\usepackage{xcolor} 
\usepackage{bigstrut}
\usepackage{tabularx}
\usepackage{upgreek}
\usepackage{subfig} 
\usepackage{epstopdf} 
\usepackage[colorlinks=true,citecolor=blue,linktoc=all]{hyperref}
  {\color{red}}%
  {}
\newcommand{\rd}{\textrm{d}}

\usepackage[normalem]{ulem}  

\begin{document}

\title{Non-equilibrium effects on stability of hybrid stars with first-order phase transitions}

\author{Peter B. Rau}
\email{prau@uw.edu}
\author{Gabriela G. Salaben}
 \affiliation{Institute for Nuclear Theory, University of Washington, Seattle, Washington 98195, USA}
 
\date{\today}

\begin{abstract}
The stability of hybrid stars with first-order phase transitions as determined by calculating fundamental radial oscillation modes is known to differ from the predictions of the widely-used Bardeen--Thorne--Meltzer criterion. We consider the effects of out-of-chemical-equilibrium physics on the radial modes and hence stability of these objects. For a barotropic equation of state, this is done by allowing the adiabatic sound speed to differ from the equilibrium sound speed. We show that doing so extends the stable branches of stellar models, allowing stars with rapid phase transitions to support stable higher-order stellar multiplets similarly to stars with multiple slow phase transitions. We also derive a new junction condition to impose on the oscillation modes at the phase transition. Termed the \textit{reactive condition}, it is physically motivated, consistent with the generalized junction conditions between two phases, and has the common rapid and slow conditions as limiting cases. Unlike the two common cases, it can only be applied to nonbarotropic stars. We apply this junction condition to hybrid stellar models generated using a two-phase equation of state consisting of nuclear matter with unpaired quark matter at high densities joined by a first-order phase transition and show that like in the slow limiting case, stars that are classically unstable are stabilized by a finite chemical reaction speed.
\end{abstract}

\maketitle

\section{Introduction}

Understanding the equation of state (EoS) of dense matter is a fundamental outstanding goal in both nuclear physics and astrophysics. One important question to improving this understanding is the nature of the phase transition between nuclear matter and deconfined quark matter. Quark deconfinement may occur via a first-order phase transition with a density discontinuity~\cite{Fukushima2011,Schmidt2017}, a second-order transition from nuclear matter to quarkyonic matter~\cite{McLerran2007,McLerran2019}, or a smooth hadron-quark crossover~\cite{Masuda2013a,Baym2019,Minamikawa2021}. Different quark phases may also exist, including color superconducting phases~\cite{Alford2019}, with associated phase transitions between them. 

Neutron stars are the principal astrophysics laboratory for studying matter at densities where the transition from hadronic to quark matter may occur, with stars containing quark matter cores termed \textit{hybrid stars}. The presence of a first-order phase transition in the dense matter EoS is of particular interest for its effect on the masses and radii of hybrid stars, since it is required for the existence of twin stars~\cite{Gerlach1968,Kampfer1981,Glendenning2000,Schertler2000,Alford2013,Benic2015,Li2020,Li2021,Alvarez-Castillo2019,Christian2021,Dexheimer2021}. These are compact stars with identical gravitational masses but different radii and internal compositions. Additional first-order phase transitions in the dense matter EoS can give rise to higher-order stellar multiplets with identical masses but different radii. In the case of only classically-stable stars-- those with mass increasing as a function of central density $\partial M/\partial\rho_c>0$-- triplet stars have been proposed~\cite{Alford2017a}. However, the presence of the phase transition modifies the definition of stellar stability away from the Bardeen--Thorne--Meltzer (BTM) criterion~\cite{Thorne1966,Bardeen1966}, part of which is that $\partial M/\partial\rho_c>0$. Taking this into account, numerous authors have demonstrated the existence of \textit{slow stable} stars~\cite{Pereira2018,Curin2021,Lugones2023}. These allow for many more pairs of twin stars consisting of a BTM-stable star plus a slow stable star.~\citet{Goncalves2022} extended the study of slow stable stars to EoS with two-phase transitions, while~\citet{Rau2023a} applied similar EoS which supported classical twin \textit{and} triplet stars to demonstrate the possible existence of higher-order slow-stable stellar multiplets-- up to six stars with identical masses but different radii.

Determining whether slow stable stars exist given an EoS requires computing the normal modes of oscillation of the star, specifically their fundamental (nodeless) radial modes. If the EoS has first-order phase transitions, junction conditions are needed to describe how the perturbation of the stellar fluid changes across the density discontinuity~\citep{Bisnovatyi-Kogan1984,Haensel1989}. The most common junction conditions are the \textit{rapid} and \textit{slow} junction conditions. Physically, these correspond to cases where the rate of the phase transition is much faster or slower than the oscillation period, such that a fluid element perturbed across the equilibrium phase boundary will either instantaneously undergo a phase transition or will retain its phase indefinitely. Stars with only rapid phase transition have identical stability properties to those determined by the BTM stability criterion, while stars with a slow phase transition do not obey this criterion in its usual form and hence may permit stable stars which would be deemed unstable according to the BTM criterion~\footnote{Though in the context of strange dwarfs-- stellar objects with typical radii and densities of white dwarfs but containing quark matter cores-- it was recently shown by~\citet{DiClemente2023} that the full BTM criterion can be applied to stars with slow phase transitions, as long as the sequence of stellar models is followed along increasing total baryon number in their cores}. 

Most studies of stability of hybrid stars with first-order phase transitions have made the simplifying assumption that the stellar fluid is always in chemical equilibrium. Since chemical equilibrium is restored through weak interactions in the bulk of the star, and the rates for these interactions are much slower than typical oscillation periods at usual temperatures for all but the youngest neutron stars, this approximation should be re-examined. Additionally, the rapid and slow phase transitions and corresponding junction conditions are purely limiting cases, and alternative conditions require considering out-of-chemical equilibrium effects. In this paper, we consider the stability of hybrid stars including non-equilibrium effects in both the bulk and at the phase transition. In the bulk, our work generalizes studies of white dwarfs~\citep{Chanmugam1977} and neutron stars~\citep{Gourgoulhon1995} without strong first-order phase transitions. At the phase transition, we introduce a novel junction condition termed the \textit{reactive} condition, showing that it generally stabilizes stars that are unstable according to the BTM criterion, but does not merely replicate the slow phase transition results. In fact, it interpolates between the slow and rapid phase transition cases, but unlike the interpolating model developed in~\citet{Rau2023a} (henceforth RS23), it is physically motivated and consistent with the generalized junction conditions introduced in~\citet{Karlovini2004}. 

In Section~\ref{sec:OutofEq} we review the calculation of radial modes of general-relativistic stars, how this is used to determine stellar stability, and discuss how non-equilibrium effects modify this calculation. The main result of this paper, the reactive junction condition, is derived in Section~\ref{sec:JunctionCondition}. Section~\ref{sec:EoS} describes the equations of state used to generate stellar models whose stability is examined when out-of-equilibrium effects are included. Section~\ref{sec:Stability} discusses the fundamental radial mode calculation with out-of-equilibrium effects and how these effects change the stability of stars with first-order phase transitions. We also discuss how the reaction mode, the radial mode which has no corresponding mode in the single-phase star, is modified using the new junction condition. Our results are reviewed in Section~\ref{sec:Conclusion}. We work in units where $c=G=\hbar=1$.

\section{Out-of-equilibrium physics and radial oscillation modes}
\label{sec:OutofEq}

Computing the radial oscillation modes of non-rotating stars in general relativity (see e.g.,~\cite{Chandrasekhar1964a,Chanmugam1977,Gondek1997}) requires first computing equilibrium stellar models for a given equation of state by solving the Tolman--Oppenheimer--Volkoff (TOV) equation. The normal modes are then found by solving two coupled first-order differential equations in the dimensionless Lagrangian displacement field $\xi$ and the Lagrangian pressure perturbation $\Delta P$, for which the angular frequency squared of the oscillation modes $\omega^2$ is an eigenvalue. This is a Sturm--Liouville problem, so stability is guaranteed if the fundamental (nodeless) mode has positive eigenvalue: $\omega_0^2>0$.

The oscillation mode equations are~\cite{Chanmugam1977,Gondek1997}:
\begin{align}
\frac{\textrm{d}\xi}{\textrm{d}r}={}&\left(\frac{\textrm{d}\nu}{\textrm{d}r}-\frac{3}{r}\right)\xi-\frac{\Delta P}{r\Gamma P},
\label{eq:dxidr}
\\
\frac{\textrm{d}\Delta P}{\textrm{d}r}={}&\Bigg[\textrm{e}^{2\lambda}\left(\omega^2 \textrm{e}^{-2\nu}-8\pi P\right)+\frac{\textrm{d}\nu}{\textrm{d}r}\left(\frac{4}{r}+\frac{\textrm{d}\nu}{\textrm{d}r}\right)\Bigg]
\nonumber
\\
{}&\times\left(\rho+P\right)r\xi
\nonumber
\\
{}&-\left[\frac{\textrm{d}\nu}{\textrm{d}r}+4\pi(\rho+P)r\textrm{e}^{2\lambda}\right]\Delta P,
\label{eq:dDeltaPdr}
\end{align}
where $P$, $\rho$ and $\Gamma$ are the pressure, energy density and polytropic index~\footnote{$\Gamma$ is often referred to as the ``adiabatic index'' in contexts where chemical equilibrium is always assumed, including in RS23. To distinguish between $\Gamma$ and the adiabatic index of the perturbation $\Gamma_1$, we employ the term ``polytropic index'' for the former in this paper, for lack of a better or commonly accepted term in the literature.} of the equilibrium background star and $r$ is the radial coordinate. $\nu$ and $\lambda$ are radial functions appearing in the metric tensor of the background star, whose first fundamental form is
\begin{equation}
\text{d}s^2 = -\text{e}^{2\nu(r)}\text{d}t^2+\text{e}^{2\lambda(r)}\text{d}r^2+r^2(\text{d}\theta^2 + \sin^2\theta \text{d}\phi^2).
\end{equation}
The formalism to compute the radial normal modes of a general relativistic star is identical to that presented in RS23: Eq.~(\ref{eq:dxidr}--\ref{eq:dDeltaPdr}) are solved subject to the boundary conditions
\begin{align}
\Delta P(r=0)={}&-3\Gamma P\xi(r=0),
\label{eq:CenterBC}
\\
\Delta P(r=R)={}&0,
\label{eq:OuterBC}
\end{align}
where $R$ is the outer radius of the star. $\xi$ is only determined up to an overall normalization factor; we take the conventional $\xi(r=0)=1$.

At a density discontinuity at a first-order phase transition, junction conditions relating the values of $\xi$ and $\Delta P$ across the discontinuity must be imposed.~\citet{Karlovini2004} found that the most general junction conditions are
\begin{subequations}
\begin{align}
\left[\xi-\mathcal{F}\right]^+_-={}&0,
\label{eq:GenJunction1}
\\
\left[(\rho+P)\mathcal{F}\right]^+_-={}&0,
\label{eq:GenJunction2}
\\
\left[\Delta P\right]^+_-={}&0,
\label{eq:GenJunction3}
\end{align}
\end{subequations}
where
\begin{equation}
\mathcal{F}=\frac{\Delta F}{r}\left(\frac{\text{d}F}{\text{d}r}\right)^{-1},
\label{eq:ScriptF}
\end{equation}
for a function $F=F(r)$ which defines the phase boundary. The subscripts refer to the high/low density ends of the phase transition. The simplest cases of junction conditions are the rapid and slow cases, corresponding to choosing $F=P$ or $\Delta F=0$ respectively. When employing the rapid junction conditions in a fundamental mode calculations, the stability as determined from $\omega_0^2>0$ matches what is found using the BTM criterion. When the slow junction conditions are used, stars which are unstable according to the BTM criterion can be stable-- these are the ``slow stable'' stars.

\subsection{Polytropic index vs. adiabatic index}

The first out-of-equilibrium effect to consider is the distinction between the polytropic index of the fluid $\Gamma$ and the adiabatic index of the perturbation $\Gamma_1$. These are defined by
\begin{equation}
\Gamma=\frac{\rho+P}{P}\frac{\textrm{d}P}{\text{d}\rho}, \qquad \Gamma_1\equiv \frac{\rho+P}{P}\left.\frac{\partial P}{\partial\rho}\right|_{s,\{Y_i\}},
\end{equation}
where the variables held constant during partial differentiation are entropy per particle $s$ and chemical species fractions $Y_i$. We work at zero temperature, so $s=0$ and only the $Y_i$ are relevant. The derivation of Eq.~(\ref{eq:dxidr}) has assumed that the fluid elements are always in chemical equilibrium with their surroundings and hence the Lagrangian perturbations of $P$ and $\rho$ are related by
\begin{equation}
\Delta P = \frac{\Gamma P}{\rho + P}\Delta\rho,
\label{eq:DeltaPEq}
\end{equation}
hence why Eq.~(\ref{eq:dxidr}) depends on $\Gamma$. If we do not assume that the fluid elements are always in chemical equilibrium with the background, we instead have
\begin{equation}
\Delta P = \frac{\Gamma_1P}{\rho+P}\Delta\rho + P\sum_i\beta_{Y_i}\Delta Y_i,
\label{eq:DeltaP}
\end{equation}
where
\begin{equation}
\beta_{Y_i} \equiv \left.\frac{\partial \ln P}{\partial Y_i}\right|_{\rho,Y_j\neq Y_i}.
\end{equation}
The standard assumption is that the Lagrangian perturbations of the species fractions are zero $\Delta Y_i=0$, which corresponds to the chemical composition of fluid elements being fixed. In this case, $\Gamma$ in Eq.~(\ref{eq:dxidr}) is replaced with $\Gamma_1$. The $r=0$ boundary condition is also changed, with $\Gamma$ in Eq.~(\ref{eq:CenterBC}) being replaced by $\Gamma_1$.

The use of the polytropic index instead of the adiabatic index is the default assumption if the EoS used to generate the equilibrium stellar models is barotropic $P=P(\rho)$. This assumption is made by most papers which have studied stability of hybrid stars with first-order phase transitions~\citep{Pereira2018,Curin2021,Goncalves2022,Lugones2023,Rau2023a}, with the justification that $\Gamma$ and $\Gamma_1$ differ by $\lesssim15$\% in the relevant range of densities in the nuclear phase~\citep{Haensel2002a}. However, in single-phase compact stars allowing the stars to be out-of-equilibrium changes the stability compared to the assumption of chemical equilibrium~\citep{Chanmugam1977,Gourgoulhon1995}, allowing some BTM-unstable stars to be stable. The effect of $\Gamma_1\neq\Gamma$ has not been examined in hybrid stars with strong first-order phase transitions prior to this work.

Given a barotropic EoS, taking $\Gamma_1\neq\Gamma$ is the only way to model out-of-equilibrium effects. The allowed values of $\Gamma_1$ are limited by causality and the Ledoux criterion for convective stability~\cite{Smeyers2010}. Since $\Gamma_1$ and $\Gamma$ are related to the adiabatic sound speed $c_s$ and equilibrium sound speed $c_{\textrm{eq}}$ by
\begin{equation}
c_s^2\equiv\frac{P\Gamma_1}{\rho+P}, \qquad c_{\textrm{eq}}^2\equiv\frac{P\Gamma}{\rho+P}, 
\end{equation}
causality $c_s,c_{\textrm{eq}}<1$ and the Ledoux criterion require that anywhere in the star
\begin{equation}
\Gamma \leq \Gamma_1 \leq 1+\frac{\rho}{P}.
\label{eq:Gamma1Condition}
\end{equation}

\subsection{The reactive junction condition}
\label{sec:JunctionCondition}

Out-of-equilibrium effects beyond taking $\Gamma_1\neq\Gamma$ require using a non-barotropic equation of state, which also allows for a new junction condition instead of the rapid and slow cases. RS23 posed the question of whether alternatives choices of $F$ in Eq.~(\ref{eq:ScriptF}) could change the stability of hybrid stars with first-order phase transitions compared to that exhibited for the rapid and slow cases. They specifically considered an intermediate speed phase transition with junction conditions
\begin{equation}
\left[\Delta P\right]^+_-=0, \qquad \left[\xi-\alpha\frac{\Delta P}{r}\left(\frac{\textrm{d}P}{\textrm{d}r}\right)^{-1}\right]^+_-=0,
\label{eq:IntermediateSpeedJunctionConditions}
\end{equation}
where the value of $0\leq\alpha\leq 1$ is varied. The slow and rapid conversion rate junction conditions are recovered when $\alpha=0$ and $\alpha=1$ respectively. For general $\alpha$ these conditions do not satisfy Eq.~\eqref{eq:GenJunction1}--\eqref{eq:GenJunction3}, an obvious weakness of this choice. However, RS23 also pointed out that for a barotropic EoS $P=P(\rho)$, $F$ can only be a function of $P$ or $r$, and any attempt to write $F$ as a more complicated function of $P$ will give $\mathcal{F}$ that is equal to the rapid conversion case. 

The obvious next step to obtain a more realistic junction condition intermediate is to allow the perturbed fluid elements to be out of chemical equilibrium. For concreteness, we consider an EoS which can be described as a function of $\rho$ and two chemical species fractions $X$ and $Y$. The EOS has a single phase transition at some density, below which we assume that $Y=0$ and $P=P(\rho,X)$, and above which $X=0$ and $P=P(\rho,Y)$. This allows us to take $F=Y$ as the definition of the phase boundary, since it drops to zero here, and hence
\begin{equation}
\mathcal{F}=\frac{\Delta Y}{r}\left(\frac{\textrm{d}Y}{\textrm{d}r}\right)^{-1}.
\label{eq:ReactiveF}
\end{equation}
Combining Eq.~(\ref{eq:GenJunction1}--\ref{eq:GenJunction2}) to eliminate $\mathcal{F}^{-}$, we obtain a junction condition for $\xi$:
\begin{equation}
[\xi]^+_-=\mathcal{F}^{+}\left(\frac{\rho^+-\rho^-}{\rho^-+P}\right).
\label{eq:ReactiveJCv1}
\end{equation}
This and Eq.~(\ref{eq:GenJunction3}) form a new set of junction conditions, but we still need to specify $\Delta Y$, which if taken to be zero simply recovers the slow junction condition.

How does allowing the perturbed fluid elements to be out of chemical equilibrium modify Eq.~(\ref{eq:dxidr}--\ref{eq:dDeltaPdr})? In the high-density phase with $P=P(\rho,Y)$, combining Eq.~(\ref{eq:DeltaP}) with
\begin{equation}
\Delta\rho = -(\rho+P)\frac{1}{r^2}\frac{\textrm{d}}{\textrm{d}r}\left(r^3\xi\right)-\frac{\textrm{d}P}{\textrm{d}r}r\xi,
\end{equation}
and \textit{not} assuming $\Delta Y=0$, we find that Eq.~(\ref{eq:dxidr}) is replaced by
\begin{equation}
\frac{\textrm{d}\xi}{\textrm{d}r}=\left(\frac{\textrm{d}\nu}{\textrm{d}r}-\frac{3}{r}\right)\xi-\frac{\Delta P}{r\Gamma_1 P}+\frac{\beta_{Y}}{r\Gamma_1}\Delta Y.
\label{eq:dxidr2}
\end{equation}
Compared to Eq.~(\ref{eq:dxidr}), Eq.~(\ref{eq:dxidr2}) replaces the equilibrium value of $\Gamma$ with the adiabatic index $\Gamma_1$, and includes an additional term $\propto \Delta Y$. In general there will be a term of this form for each chemical species fraction in a given phase of matter.

Similar to Eq.~(\ref{eq:DeltaP}), for the Eulerian perturbation of $\rho$, using $\delta Y = \Delta Y - r\xi (\textrm{d}Y/\textrm{d}r)$ for purely radial motion,
\begin{equation}
\delta P = \frac{\Gamma_1P}{\rho+P}\delta\rho + \xi P\beta_{Y}\frac{\textrm{d}Y}{\textrm{d}r} - P\beta_{Y}\Delta Y.
\end{equation}
Using this in the derivation of the normal mode equation for $\textrm{d}\Delta P/\textrm{d}r$ in e.g.,~\cite{Chandrasekhar1964a}, we obtain
\begin{align}
\frac{\textrm{d}\Delta P}{\textrm{d}r}={}&\Bigg[\textrm{e}^{2\lambda}\left(\omega^2 \textrm{e}^{-2\nu}-8\pi P\right)+\frac{\textrm{d}\nu}{\textrm{d}r}\left(\frac{4}{r}+\frac{\textrm{d}\nu}{\textrm{d}r}\right)
\nonumber
\\
{}& \quad -\frac{\beta_{Y}}{2\Gamma_1}\frac{\textrm{d}\nu}{\textrm{d}r}\frac{\textrm{d}Y}{\textrm{d}r}\Bigg]\left(\rho+P\right)r\xi
\nonumber
\\
{}&-\left[\frac{\textrm{d}\nu}{\textrm{d}r}+4\pi(\rho+P)r\textrm{e}^{2\lambda}\right]\Delta P.
\label{eq:dDeltaPdr2}
\end{align}
We see that compared to Eq.~(\ref{eq:dDeltaPdr}) of RS23 there is an additional term in the coefficient of $\xi$ proportional to the gradient of $Y$, and that this term should be included even when $\Delta Y=0$. This term is proportional to the Brunt--V\"{a}is\"{a}l\"{a} frequency squared $N^2$ term that appears in the nonradial oscillation mode equations~\citep{McDermott1983}, and could be included instead by replacing $\omega^2$ with $\omega^2-N^2$ in the first term on the right-hand side of Eq.~(\ref{eq:dDeltaPdr2}). However, no purely radial $g$-modes exist, and the effect of this term is to shift the radial mode frequencies.

Starting with the equation for total baryon conservation $\nabla_{\mu}n^{\mu}_b=0$
for baryon four-current $n^{\mu}_b$, and assuming that we can write similar equations for the different fluid species in each phase, we obtain an equation describing the evolution of the chemical fractions. For fraction $Y$, this is
\begin{equation}
\textrm{e}^{-\nu/2}\gamma(|v^r|)\left(\frac{\partial Y}{\partial t} + v^r\frac{\partial Y}{\partial r}\right)=\frac{\gamma_{Y}}{n_b},
\end{equation}
where $\gamma_Y$ is the volumetric creation rate of the particles with chemical fraction $Y$, $\gamma(|v^r|)$ is the Lorentz factor and $v^r$ is the radial velocity (assuming radial motion only) of the fluid. $n_b$ is the baryon number density. Taking the Lagrangian perturbation of this and retaining only terms to lowest order in the velocity gives
\begin{equation}
\textrm{e}^{-\nu/2}\frac{\partial\Delta Y}{\partial t}=\frac{\gamma_{Y}}{n_b^2}(Q_b-1)\Delta n_b + \frac{\gamma_{Y}}{n_b}Q_{Y}\Delta Y,
\end{equation}
where
\begin{equation}
Q_b\equiv\frac{\partial \ln\gamma_{Y}}{\partial\ln n_b}, \quad Q_{Y}\equiv\frac{\partial \ln\gamma_{Y}}{\partial Y}.
\end{equation}
Assuming harmonic time dependence $\Delta Y\propto e^{-i\omega t}$ where $\omega$ is the oscillation angular frequency, this equation can be rearranged as
\begin{equation}
\Delta Y =\frac{\gamma_{Y}(1-Q_b)\left(\gamma_{Y}Q_{Y}-i\omega \textrm{e}^{-\nu/2}n_b\right)}{(\gamma_{Y}Q_{Y})^2+\omega^2\textrm{e}^{-\nu}n_b^2}\frac{\Delta n_b}{n_b}.
\label{eq:DeltaY}
\end{equation}
In the limit that the reactions are much faster than the oscillation period, we can take $\omega\rightarrow 0$ in Eq.~(\ref{eq:DeltaY}) and obtain
\begin{equation}
\Delta Y\approx -\frac{Q_b-1}{Q_{Y_s}}\frac{\Delta n_b}{n_b}\equiv -Z\frac{\Delta n_b}{n_b},
\label{eq:DeltaY2}
\end{equation}
where $Z$ is a parameter characterizing the rate of restoration of equilibrium. It can in principle be calculated given a microscopically-computed $\gamma_{Y}$. Not making the assumption $\omega\rightarrow 0$ in Eq.~(\ref{eq:DeltaY}) gives rise to dissipation in the form of bulk viscosity, which makes the $\omega$ values complex and violates the assumptions of Sturm--Liouville theory~\cite{Weinberg1972}.

The weak reactions that restore equilibrium in the bulk of the star are much slower than typical oscillation period except at very high temperatures~\citep{Friman1979}. Hence we can take $\Delta Y\approx 0$ in the bulk of the star to very good approximation, and can ignore the $\Delta Y$ term in Eq.~(\ref{eq:dxidr2}) and~(\ref{eq:dDeltaPdr2}). But the reactions that occur at the phase transition, including quark (de)confinement, could be faster than the oscillation period. So we should retain $\Delta Y$ near the phase transition. For a nuclear matter to deconfined quark matter phase transition, $Y$ is the strange quark fraction, which is zero in nuclear matter and nonzero in quark matter.

Combining Eq.~(\ref{eq:DeltaP}),~(\ref{eq:DeltaY2}), and
\begin{equation}
\frac{\Delta\rho}{\rho+P}=\frac{\Delta n_b}{n_b},
\label{eq:DeltarhoDeltan}
\end{equation}
we obtain
\begin{equation}
\Delta Y = -\frac{Z}{\Gamma_1-\beta_{Y_s} Z}\frac{\Delta P}{P}.
\label{eq:DeltaY3}
\end{equation}
Inserting this into Eq.~(\ref{eq:ReactiveF}) and then eliminating $\mathcal{F}^{+}$ from Eq.~(\ref{eq:ReactiveJCv1}) gives
\begin{equation}
\left[\xi\right]^+_-=\frac{-Z}{\Gamma^+_1-\beta^+_{Y}Z}\frac{\Delta P}{rP}\left(\frac{\textrm{d}Y}{\textrm{d}r}\right)^{-1}_+\left(\frac{\rho^--\rho^+}{\rho^-+P}\right).
\label{eq:ReactiveJC}
\end{equation}
We have used that $\Delta P$, $P$ and $r$ are continuous across the junction. This equation and Eq.~(\ref{eq:GenJunction3}) form a new set of junction conditions, which we term the \textit{reactive} conditions. 

The reactive junction conditions are a physically-motivated set of junction conditions that are consistent with the generalized junction conditions Eq.~(\ref{eq:GenJunction1})--(\ref{eq:GenJunction3}) and which interpolate between the rapid and slow conditions. First, in the limit $Z\rightarrow 0$, Eq.~(\ref{eq:ReactiveJC}) clearly reduces to $\left[\xi\right]^+_-=0$, the junction condition for $\xi$ in the slow case. To recover the rapid case, we note that
\begin{align}
\frac{\rd P}{\rd r} ={}& \frac{\Gamma_1P}{n_b}\frac{\rd n_b}{\rd r} + P\beta_Y\frac{\rd Y}{\rd r}
\nonumber
\\
={}&P\left(\Gamma_1\left(\frac{\rd Y}{\rd\ln n_b}\right)^{-1}\hspace{-2.5mm} + \beta_Y\right)\frac{\rd Y}{\rd r}.
\end{align}
Using this to eliminate $\beta_Y$ from Eq.~(\ref{eq:ReactiveJC}) gives
\begin{align*}
\left[\xi\right]^+_-= \frac{-Z\frac{\Delta P}{rP}\left(\frac{\rho^--\rho^+}{\rho^-+P}\right)}{\Gamma_1^+\left[1+Z\left(\frac{\rd Y}{\rd\ln n_b}\right)^{-1}_+\right]\left(\frac{\rd Y}{\rd r}\right)_+-Z\left(\frac{\rd\ln P}{\rd r}\right)_+}.
\end{align*}
In the case $Z\rightarrow-(\rd Y/\rd\ln n_b)_+$, Eq.~(\ref{eq:ReactiveJC}) becomes
\begin{align*}
\left[\xi\right]^+_-={}&\frac{\Delta P}{r}\left(\frac{\textrm{d}P}{\textrm{d}r}\right)_+^{-1}\left(1-\frac{\rho^++P}{\rho^-+P}\right)
\nonumber
\\
={}&\frac{\Delta P}{r}(\rho^++P)\left(\frac{\textrm{d}P}{\textrm{d}r}\right)_+^{-1}\left(\frac{1}{\rho^++P}-\frac{1}{\rho^-+P}\right)
\nonumber
\\
={}&\left[\frac{\Delta P}{r}\left(\frac{\textrm{d}P}{\textrm{d}r}\right)^{-1}\right]_-^+,
\end{align*}
which is the junction condition on $\xi$ in the rapid case (the second equation in Eq.~(\ref{eq:IntermediateSpeedJunctionConditions}) with $\alpha=1$). To show this we used from the TOV equation that
\begin{equation}
\frac{1}{\rho+P}\frac{\textrm{d}P}{\textrm{d}r}=-\frac{m+4\pi r^3P}{r^2(1-2m/r)},
\end{equation}
is continuous across the phase transition because $P$, $r$ and the enclosed mass $m=m(r)$ are continuous across the phase transition. $Z=-(\rd Y/\rd\ln n_b)_+$ being the rapid limit is expected because this says that the relation between the perturbations of $Y$ and $n_b$ as given by Eq.~(\ref{eq:DeltaY2}) is identical to its value in chemical equilibrium, and the rapid case assumes that the fluid elements are always in chemical equilibrium.

The range of physically meaningful values of $Z$ is $-(\rd Y/\rd\ln n_b)_+\leq Z\leq 0$. This is because as the reaction rate increases, $Z$ decreases from $0$ to $-(\rd Y/\rd\ln n_b)_+$. $Z>0$ would correspond to a slower reaction rate than infinitely slow and is unphysical, and $Z<-(\rd Y/\rd\ln n_b)_+$ would be a faster reaction rate than infinitely fast. We note that permitting small values of $Z$ which imply slow phase-changing reactions is inconsistent with the assumption that underlined the derivation of the reactive junction conditions. However, we will see that values of $Z$ within the same order of magnitude as the rapid-case recovering $Z=-(\rd Y/\rd\ln n_b)_+$ still give rise to fundamentally different behavior than the rapid case and stabilize stars similar to what occurs in the slow case.

In deriving Eq.~(\ref{eq:ReactiveJC}), we take a linear combination of Eq.~(\ref{eq:GenJunction1}--\ref{eq:GenJunction2}) to eliminate $\mathcal{F}^-$. This is not a unique choice and we could alternatively eliminate $\mathcal{F}^+$ from the equations, which would give a similar boundary condition to Eq.~(\ref{eq:ReactiveJC}) but one which depends on a chemical species fraction and its gradient on the lower density side of the phase transition. This boundary condition would depend on a different parameter characterizing the equilibration rate in the low density phase which we call $Z^-$, and the physically allowed range of $Z^{-}$ would be different from the allowed range of $Z$. If we choose to satisfy Eq.~(\ref{eq:GenJunction1}--\ref{eq:GenJunction2}) independently, the value of either $Z^{-}$ or $Z$ would be constrained, leaving the other as free parameter (which should in principle be a quantity that can be computed). That $Z$ and $Z^{-}$ must be related to each other this way is not surprising since the restoration of equilibrium involves reactions depending on the chemical species fractions on both sides of the transition.

\section{Equations of state}
\label{sec:EoS}

\subsection{Three-phase barotropic EoS}
\label{sec:EoS1}

\begin{figure}
\includegraphics[width=\columnwidth]{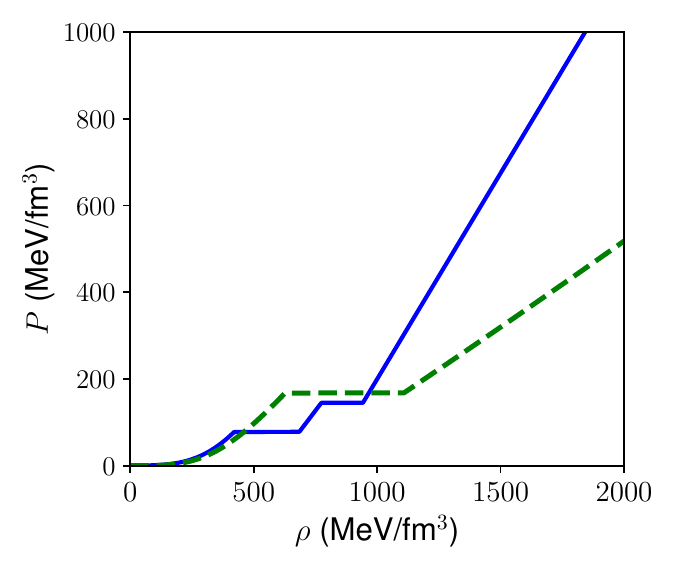}
\caption{Equations of state used in this paper: three-phase EoS with nuclear phase and two quark phases with constant equilibrium sound speeds as described in Section~\ref{sec:EoS1} (solid line), and the two-phase EoS with nuclear plus quark phase as described in Section~\ref{sec:EoS2} (dashed line).}
\label{fig:EoSs}
\end{figure}

To study slow-stable stars with high-order stellar multiplets, RS23 used the three-phase, chemically equilibrated (i.e, barotropic) EoS with constant equilibrium sound speeds for the quark phases described in Ref.~\cite{Alford2017a}. The low-density nuclear phase is joined to two color-superconducting quark phases (termed 2SC and CFL respectively) by the Maxwell construction with large density discontinuities. We use one parametrization of this EoS given in Table~\ref{tab:3PParameters} to study the simplest non-equilibrium effects on stability with a barotropic EoS. The definition of the EoS parameters matches~\citet{Alford2017a} or RS23. This parametrization supports BTM-stable triplet stars, and stellar masses up to $1.88M_{\odot}$, consistent with the neutron star maximum mass constraints within $3\sigma$ confidence intervals~\cite{Demorest2010,Antoniadis2013,Cromartie2020,Abbott2017a,Riley2019,Miller2019,Riley2021,Miller2021}. The EoS is plotted in FIG.~(\ref{fig:EoSs}).

\begin{table}
	\caption{Parametrization for the nuclear plus two quark phase EoS used in this paper. $\rho_1$ and $\rho_2$ are the energy densities at the low end of the nuclear-2SC and 2SC-CFL phase transitions, $P_1$ is the pressure at the nuclear-2SC phase transition, $\Delta\rho_1$ and $\Delta\rho_2$ are the energy density discontinuities of the nuclear-2SC and 2SC-CFL phase transitions; all are given in MeV/fm$^3$. $c_{\textrm{eq}1}^2$ and $c_{\textrm{eq}2}^2$ are the equilibrium sound speeds squared in the 2SC and CFL quark phases in units of $c^2$.}
	  \centering
    \begin{tabular}{|c|c|c|c|c|c|c|}
    \hline
	\multicolumn{1}{|c|}{$\rho_1$} & $P_1$ & $\rho_2$ & $\Delta\rho_1$ & $\Delta\rho_2$ & $c_{\textrm{eq}1}^2$ & $c_{\textrm{eq}2}^2$\\ 
    \hline
    420.7 & 77.7 & 774.1 & 263.6 & 168.3 & 0.75 & 0.95 \\
	\hline   
    \end{tabular}
    \label{tab:3PParameters}
\end{table}

The non-equilibrium effects are modeled by choosing $\Gamma_1\neq\Gamma$. Since the nuclear phase of the EoS is based on a tabulated model, $\Gamma$ here is derived from finite differencing $P(\rho)$. For simplicity, we let $\Gamma_1=\Gamma$ in this region and restrict our study of non-equilibrium effects to the two quark phases. We do this by specifying constant values of $c^2_s$ in the two quark phases subject to Eq.~(\ref{eq:Gamma1Condition}) i.e., $c_{\textrm{eq}1}^2<c_{s1}^2<1$ and $c_{\textrm{eq}2}^2<c_{s2}^2<1$. The different choices of the $c_{s}^2$ values we examine are listed in Table~\ref{tab:3PCs2ad}.

\begin{table}
	\caption{Choices of constant adiabatic sound speeds squared $c_{s}^2$ in units of $c^2$ for the 2SC phase ($c_{s1}^2$) and CFL phase ($c_{s2}^2$). Note configuration 1 is equivalent to choosing $c_{s}^2=c_{\textrm{eq}}^2$ in both phases.}
	  \centering
    \begin{tabular}{|c|c|c|}
    \hline
	\multicolumn{1}{|c|}{Name} & $c_{s1}^2$ & $c_{s2}^2$\\ 
    \hline
    	1 & 0.75 & 0.95\\
	\hline  
	    2 & 0.8 & 0.95\\
	\hline 
	    3 & 0.9 & 0.95\\
	\hline 
	    4 & 0.8 & 1\\
	\hline 
	    5 & 0.9 & 1\\
	\hline 
	    6 & 1 & 1\\
	\hline 
    \end{tabular}
    \label{tab:3PCs2ad}
\end{table}

\subsection{Two-phase nonbarotropic EoS}
\label{sec:EoS2}

When applying the reactive junction condition, an EoS which includes information about chemical species fractions must be used. For this purpose we use a composite EoS with a single first-order phase transition between nuclear matter and deconfined, unpaired, three-flavor quark matter. The crust is taken from the DDME2 EoS~\cite{Colucci2013}, which is joined continuously to the Zhao--Lattimer EoS~\citep{Zhao2020} in the form used in Ref.~\citep{Constantinou2021} for the nuclear (neutron-proton-electron matter) phase, though we do not include muons. The quark phase EoS is based on the vMIT model as used in Refs.~\citep{Zhao2020,Constantinou2021}, but joined to the nuclear matter phase using a Maxwell construction. We use the nuclear phase parameters given by EoS XOA in Ref.~\citep{Constantinou2021}.  Our treatment of the quark matter phase differs somewhat from the references and so is discussed in detail below. We also describe the calculation of $\Gamma_1$ and the $\beta_Y$.

\begin{table}
\caption{Parametrization of quark matter EoS used in this paper, and properties of the resulting first-order phase transition using the Maxwell construction between nuclear and quark matter. $m_s$ is the strange quark mass, $B$ is the bag constant, $a_V$ is the vector meson coupling constant defined in Eq.~(\ref{eq:a_V}), $P_Q$ is the phase transition pressure, $\rho_Q$ is the energy density at the low-density end of the phase transition, $\Delta\rho_Q$ is the energy density discontinuity across the phase transition. $m_s$, $B$ and $a_V$ differ from those values chosen in Ref.~\citep{Constantinou2021}.}
	  \centering
    \begin{tabular}{|c|c|c|c|c|}
    \hline
	\multicolumn{1}{|c|}{Parameter} & Value \\ 
    \hline
    $m_s$ & 100 MeV \\
	\hline
	$B$ & (190 MeV)$^4$ \\
	\hline  
	$a_V$ & $1.541\times10^{-5}$ MeV$^{-2}$ \\
	\hline  
	$P_Q$ & 167.6 MeV/fm$^3$ \\
	\hline  
	$\rho_Q$ & 627.2 MeV/fm$^3$ \\
	\hline  
	$\Delta\rho_Q$ & 480.2 MeV/fm$^3$\\
	\hline  
    \end{tabular}
    \label{tab:UQMParameters}
\end{table}

In the quark matter phase, we start with a relativistic mean-field model with pressure
\begin{equation}
P = \sum_q P_q + P_e - B +\frac{1}{2}m_V^2V^2,
\end{equation}
where the $P_q$ are the quark pressure contributions $q=\{u,d,s\}$, $P_e$ is the electron pressure contribution, $B$ is the bag constant and $V$ is the vector meson field with mass $m_V$. We assume massless up and down quarks $m_u=m_d=0$ and electrons $m_e=0$, and ignore muons. $P_q$ and $P_e$ are
\begin{subequations}
\begin{align}
P_q ={}& \frac{\mu_q^{*4}}{4\pi^2}, \quad q=u,d,
\\
P_s ={}& \frac{1}{8\pi^2}\Bigg[\mu_s^*(2\mu_s^{*2}-5m_s^2)\sqrt{\mu_s^{*2}-m_s^2}
\nonumber
\\
{}&\qquad +3m_s^4\ln\left(\frac{\mu_s^*+\sqrt{\mu_s^{*2}-m_s^2}}{m_s}\right)\Bigg], 
\\
P_e={}&\frac{\mu_e^4}{12\pi^2},
\end{align}
\end{subequations}
for strange quark mass $m_s$, electron chemical potential $\mu_e$ and where
\begin{equation}
\mu_q^* = \mu_q - g_{V}V,
\end{equation}
is the effective chemical potential for each quark species for vector meson coupling constant $g_{V}$. $\mu_q$ is the bare quark chemical potential. The number densities are then
\begin{subequations}
\begin{align}
n_q={}&\left.\frac{\partial P}{\partial\mu_q}\right|_{\mu_q'\neq\mu_q,V}=\frac{\partial P}{\partial\mu_q^*}\left.\frac{\partial\mu_q^*}{\partial\mu_q}\right|_{V}=\frac{\mu_q^{*3}}{\pi^2}, \quad q=u,d,
\label{eq:n_q}
\\
n_s={}&\frac{(\mu_s^{*2}-m_s^2)^{3/2}}{\pi^2},
\label{eq:n_s}
\\
n_e={}&\frac{\mu_e^{3}}{3\pi^2}.
\end{align}
\end{subequations}
We determine the value of $V$ by maximizing the pressure with respect to it:
\begin{equation}
\frac{\partial P}{\partial V} = 0 = \sum_{q}\frac{\partial P}{\partial \mu^*_q}\left.\frac{\partial \mu^*_q}{\partial V}\right|_{\mu_q}+m_{V}^2V=-g_{V}\sum_qn_q+m_{V}^2V.
\end{equation} 
Since $\sum_qn_q=3n_b$, in equilibrium we find
\begin{equation}
V = 3\left(\frac{g_{V}}{m^2_{V}}\right)n_b.
\label{eq:VFieldEquilibrium}
\end{equation}

Electrical charge neutrality and weak equilibrium require that
\begin{subequations}
\begin{align}
\frac{2}{3}n_u ={}& \frac{1}{3}(n_d+n_s) + n_e.
\label{eq:ChargeNeutrality}
\\
\mu_u ={}& \mu_d + \mu_e,
\label{eq:WeakEquilibrium1}
\\
\mu_s ={}& \mu_d.
\label{eq:WeakEquilibrium2}
\end{align}
\end{subequations}
Requiring that Eq.~(\ref{eq:ChargeNeutrality}--\ref{eq:WeakEquilibrium2}) simultaneously allows us to calculate the chemical potentials and number densities in equilibrium. The energy density $\rho$ is
\begin{equation}
\rho = \sum_i\left.\frac{\partial P}{\partial\mu_i}\right|_{V}\mu_i-P=\sum_in_i\mu^*_i+n_e\mu_e+a_Vn_b^2-P.
\label{eq:EnergyDensity}
\end{equation}
where we have inserted Eq.~(\ref{eq:VFieldEquilibrium}) and defined
\begin{equation}
a_V\equiv \left(\frac{3g_{V}}{m_{V}}\right)^2.
\label{eq:a_V}
\end{equation}
The quark matter EoS parametrization is given in Table~\ref{tab:UQMParameters}. We choose different parameters from those in Ref.~\citep{Constantinou2021}: this difference arises from requiring the astrophysical constraint of a $2M_{\odot}$ star be met while having a first-order phase transition, whereas Ref.~\citep{Constantinou2021} considers a crossover phase transition. The pressure $P_Q$, energy density at lower end of phase transition $\rho_Q$ and energy density discontinuity at the phase transition $\Delta\rho_Q$ are also given in this table. This combined crust-nuclear-quark matter EoS supports a $>2M_{\odot}$ hybrid star, and is plotted in FIG.~\ref{fig:EoSs}. We have checked that choosing EoS parameters such that there is no stable hybrid star branch does not qualitatively change our findings.

To compute the adiabatic index, we express $P$ as a function of $\rho$ and the various particle species fractions. The total baryon number density is given in terms of the quark number densities by
\begin{equation}
n_b = \frac{1}{3}(n_u+n_d+n_s).
\end{equation}
Define quark flavor fractions $Y_q\equiv n_q/(3n_b)$, and $Y_e\equiv n_e/n_b$. Using $1=Y_u+Y_d+Y_s$ and Eq.~(\ref{eq:ChargeNeutrality}), we find
\begin{equation}
Y_u = \frac{1+Y_e}{3}, \qquad Y_d = \frac{2-Y_e}{3} - Y_s,
\end{equation}
which allows us to express all quark and electron number densities in terms of $n_b$, $Y_e$ and $Y_s$ only. Doing so, the resulting adiabatic index is
\begin{equation}
\Gamma_1=\frac{\rho+P}{P}\left.\frac{\partial P}{\partial \rho}\right|_{Y_e,Y_s},
\end{equation}
where
\begin{align}
\left.\frac{\partial P}{\partial \rho}\right|_{Y_e,Y_s}=\frac{1}{\mu_n}\Bigg[{}&\left(1+Y_e\right)n_u\left(\frac{\partial n_u}{\partial \mu_u^*}\right)^{-1}
\nonumber
\\
{}&+\left(2-Y_e-3Y_s\right)n_d\left(\frac{\partial n_d}{\partial \mu^*_d}\right)^{-1}
\nonumber
\\
{}&+3Y_sn_s\left(\frac{\partial n_s}{\partial \mu_s^*}\right)^{-1}+Y_en_e\left(\frac{\partial n_e}{\partial \mu_e}\right)^{-1}\Bigg]
\nonumber
\\
{}& + \frac{a_Vn_b}{\mu_n}.
\end{align}
where $\mu_n=\mu_u+2\mu_d$ is the neutron chemical potential (note that the bare quark chemical potentials appear here). The partial derivatives of the number densities are readily computed from Eq.~(\ref{eq:n_q}--\ref{eq:n_s}). To compute $\beta_{Y_s}$ and $\beta_{Y_e}$ we also need
\begin{align}
\left.\frac{\partial P}{\partial Y_s}\right|_{\rho,Y_e}={}&3n_b\left[n_s\left(\frac{\partial n_s}{\partial \mu_s^*}\right)^{-1}-n_d\left(\frac{\partial n_d}{\partial \mu_d^*}\right)^{-1}\right],
\\
\left.\frac{\partial P}{\partial Y_e}\right|_{\rho,Y_s}={}&n_b\Bigg[n_u\left(\frac{\partial n_u}{\partial \mu_u^*}\right)^{-1}+n_e\left(\frac{\partial n_e}{\partial \mu_e}\right)^{-1}
\nonumber
\\
{}&\qquad-n_d\left(\frac{\partial n_d}{\partial \mu_d^*}\right)^{-1}\Bigg].
\label{eq:dPdY_e}
\end{align}
Eq.~(\ref{eq:dPdY_e}) evaluates to zero for massless $u$ and $d$ quarks and electrons in weak equilibrium. A similar procedure is done for the nuclear phase, though with only one relevant chemical species fraction, the proton fraction $Y=n_p/n_b$. Because of the $\propto \textrm{d}Y/\textrm{d}r$ term in Eq.~(\ref{eq:dDeltaPdr2}), the chemical fractions need to be computed in the entire background star and not simply at the nuclear-quark phase transition as needed to use the reactive junction condition.

\section{Effects on stellar stability}
\label{sec:Stability}

\subsection{Three-phase barotropic EoS}

\begin{figure}
\includegraphics[width=\columnwidth]{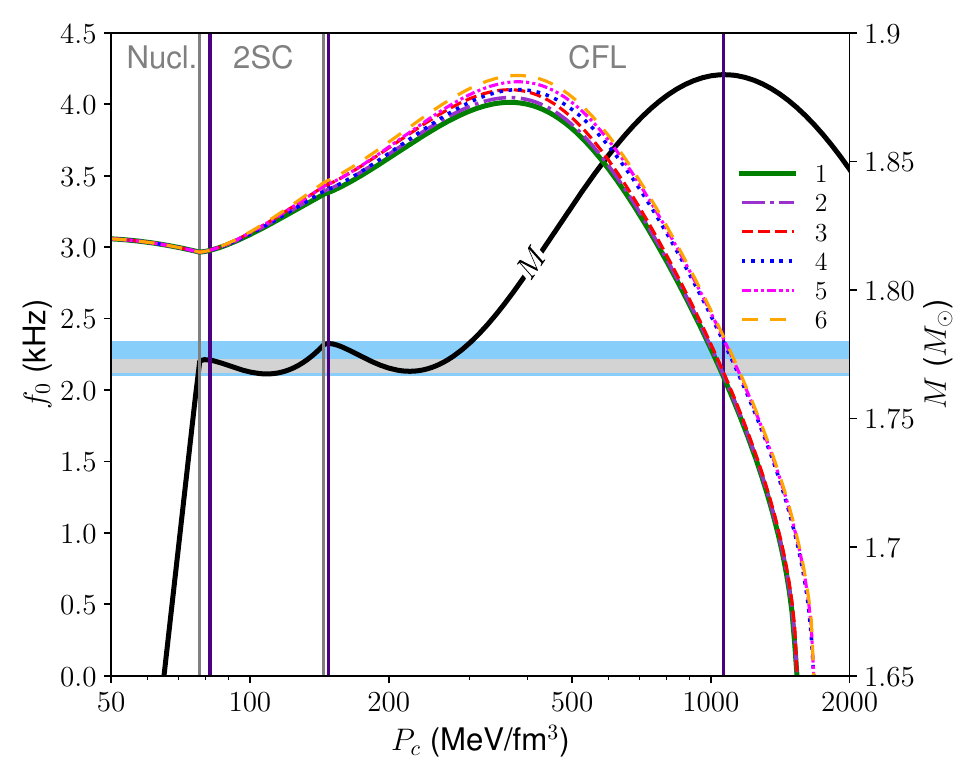}
\caption{Fundamental radial mode frequency $f_0$ (left) and stellar mass $M$ (right) as functions of central pressure $P_c$ using the three-phase barotropic EoS of Section~\ref{sec:EoS1} with the slow-slow phase transition configuration. $f_0<0$ corresponds to an imaginary (unstable) frequency. Non-equilibrium effects are included by taking different adiabatic sound speed squared $c_s^2$ than the equilibrium sound speed squared $c_{\text{eq}}^2$. The different curves correspond to the frequencies found for the choices of $c_s^2$ described in Table~\ref{tab:3PCs2ad}. Solid vertical lines are placed at the central pressures corresponding to the phase transitions (gray) and the local maxima in $M$ (indigo). The shaded gray horizontal band of stellar masses supports quintuplet stars, while the shaded blue bands support triplet stars. The phase in the center of the star is labeled at the top of the plot, with the phases separated by the vertical gray lines.}
\label{fig:3PModesSS}
\end{figure}

\begin{figure}
\includegraphics[width=\columnwidth]{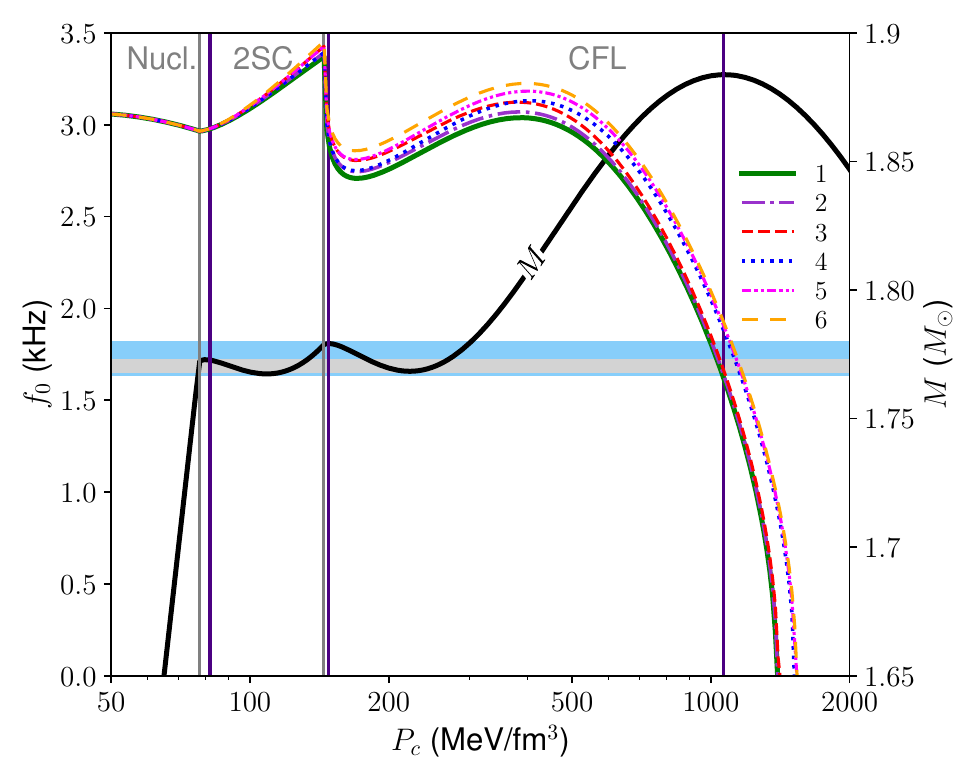}
\caption{Same as FIG.~\ref{fig:3PModesSS} except for a slow-rapid phase transition configuration.}
\label{fig:3PModesSR}
\end{figure}

\begin{figure}
\includegraphics[width=\columnwidth]{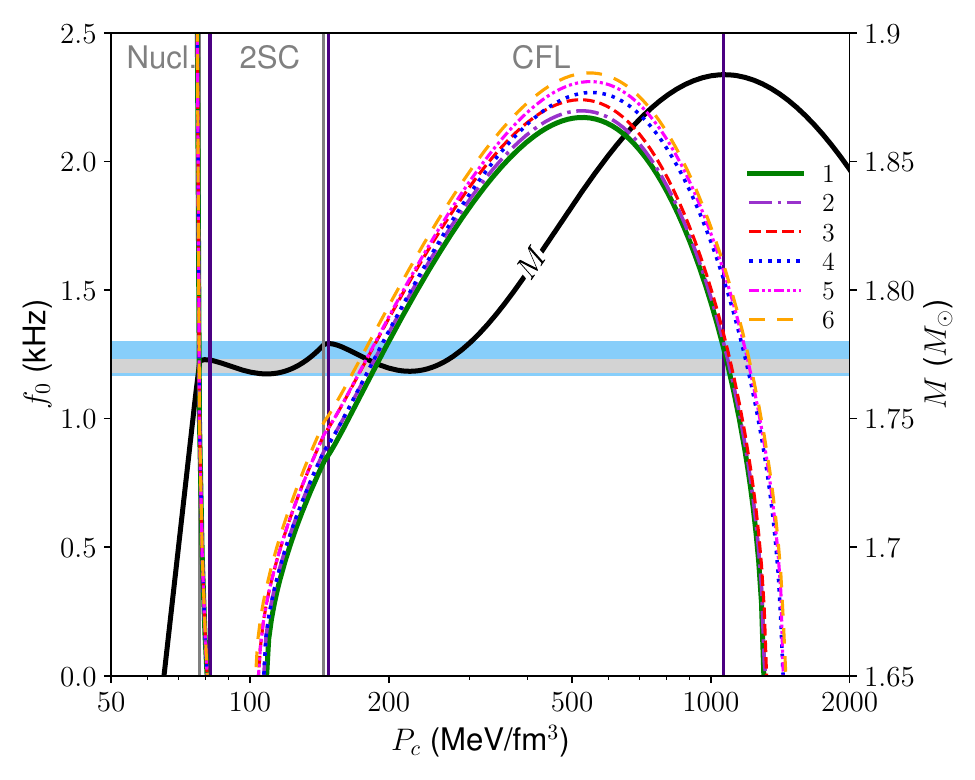}
\caption{Same as FIG.~\ref{fig:3PModesSS} except for a rapid-slow phase transition configuration.}
\label{fig:3PModesRS}
\end{figure}

\begin{figure}
\includegraphics[width=\columnwidth]{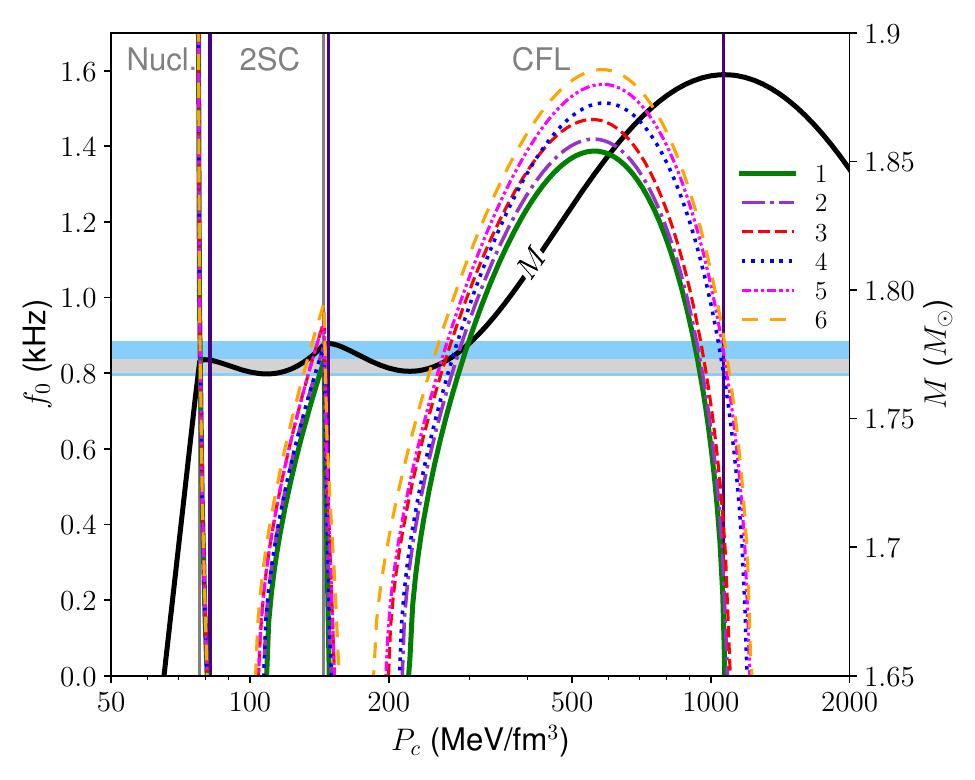}
\caption{Same as FIG.~\ref{fig:3PModesSS} except for a rapid-rapid phase transition configuration.}
\label{fig:3PModesRR}
\end{figure}

We solved Eq.~(\ref{eq:dxidr}--\ref{eq:dDeltaPdr}) subject to boundary conditions Eq.~(\ref{eq:CenterBC}--\ref{eq:OuterBC}) and with $\Gamma\rightarrow\Gamma_1$ for stellar models constructed with the EoS presented in Section~\ref{sec:EoS1}. Slow and rapid junction conditions as described following Eq.~(\ref{eq:ScriptF}) were imposed at the phase transitions. The calculation was performed using the shooting method. FIG.~\ref{fig:3PModesSS}--\ref{fig:3PModesRR} shows the fundamental radial mode frequency $f_0=\omega_0/(2\pi)$ as a function of stellar central pressure $P_c$ for this EoS and the six different choices of $c^2_s$ in Table~\ref{tab:3PCs2ad}. The four different permutations of slow and rapid junction conditions at the two phase transitions are considered; the configurations are referenced with the phase transition rate at the lower density transition (nuclear-2SC) first. To show more detail of the $f_0$ curves, the $f_0$ values within the nuclear phase are only shown for the slow-slow and slow-rapid configurations (FIG.~\ref{fig:3PModesSS}--\ref{fig:3PModesSR}). The slow-slow and slow-rapid configurations both clearly support slow-stable quintuplets consisting of three BTM-stable stars and two slow-stable stars within the gray shaded band around $M=1.77M_{\odot}$, and two triplets consisting of two BTM-stable and one slow-stable star at slightly higher and lower masses shaded light blue. In chemical equilibrium (case 1), the rapid-slow configuration supports slow-stable quadruplet stars, while the rapid-rapid configuration only supports triplet stars.

FIG.~\ref{fig:3PModesSS}--\ref{fig:3PModesRR} clearly show that increasing $\Gamma_1$ above $\Gamma$ (increasing $c_s^2$ above $c^2_{\textrm{eq}}$) expands the range of central pressures for which $f_0>0$ and the hybrid stars are stable. This holds for all permutations of the junction conditions.  This is consistent with findings~\citep{Chanmugam1977,Gourgoulhon1995} for single-phase stars which considered $\Gamma_1\neq\Gamma$, where increasing $\Gamma_1$ infinitesimally above $\Gamma$ stabilizes stars that were BTM-unstable. The further above $\Gamma$ that $\Gamma_1$ is, the greater the range of central pressures that are stabilized beyond those already supporting a stable star using the fully equilibrated model. One of the most noticeable effects, which is clear for all permutations of the junction conditions, is that as $P_c$ increases, the influence of the CFL phase increases and the $f_0$ values converge into two nearly-overlapping curves, one for configurations 1--3 ($c_{s2}^2=0.95$) and one for configurations 4--6 ($c_{s2}^2=1$). 

\begin{figure}
\includegraphics[width=\columnwidth]{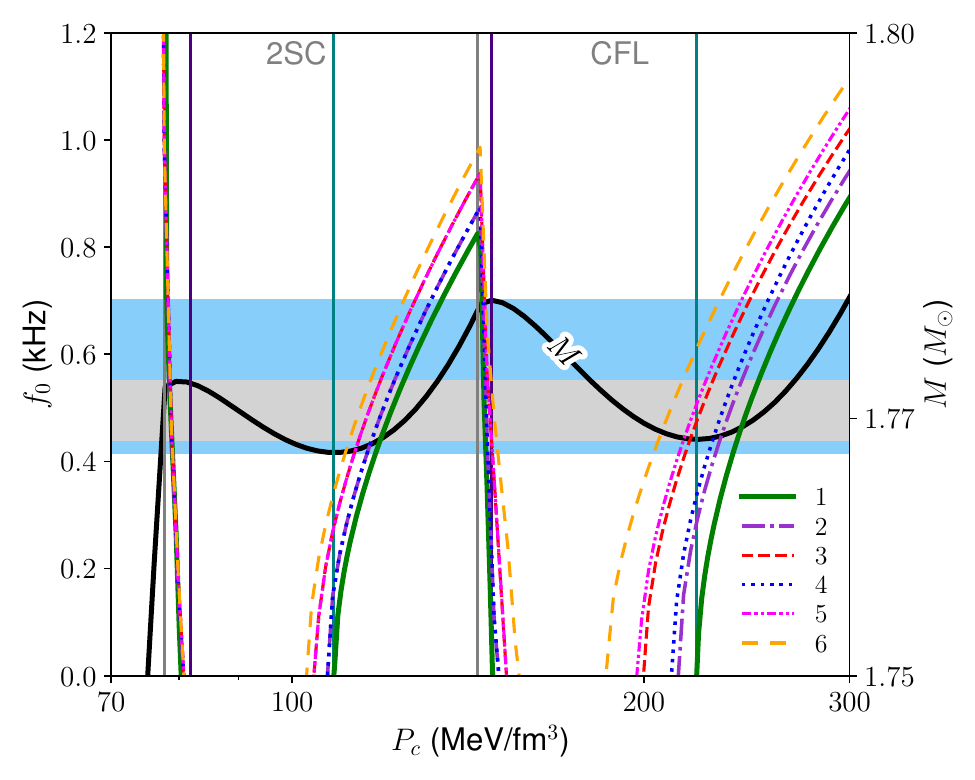}
\caption{Same as FIG.~\ref{fig:3PModesRR} except zoomed in to show detail around the first two local maxima in $M$. The local minima in $M$ are indicated with vertical teal lines.}
\label{fig:3PModesRRZoom}
\end{figure}

The main effect of $\Gamma_1\neq\Gamma$ at lower $P_c$ is expanding the stable range of $P_c$ to lower values than those corresponding to the local minima in $M$, which allows the rapid-rapid and rapid-slow configurations to also support a higher-order stellar multiplet that the BTM-stable triplet stars like in the slow-slow and slow-rapid cases. For this EoS, the rapid-rapid and rapid-slow configurations with $\Gamma_1\neq\Gamma$ support an additional stable star below the $P_c=223$ MeV/fm$^3$ local minimum in $M$, as shown in greater detail in FIG.~\ref{fig:3PModesRRZoom}.  However, the chosen values of $c_s^2$ are unable to stabilize to sufficiently low $P_c$ below the $P_c=110$ MeV/fm$^3$ local minimum in $M$ to support stars with masses in the quintuplet band at $P_c$ between this local minimum and the $P_c=82$ MeV/fm$^3$ local maximum in $M$. Thus the rapid-rapid and rapid-slow configurations only support stable quadruplet stars. The additional range of stable $P_c$ values above the local maximum in $M$ at $P_c=148$ MeV/fm$^3$ does allow a separate stable triplet of stars in the rapid-rapid configuration, matching a similar stable triplet which is present for the other three configurations at masses just above that for which the slow-slow and slow-rapid configurations support quintuplet stars. These observations show that non-equilibrium effects can mimic the slow-stabilization effects through by extending the range of $P_c$ which correspond to stable objects, though the differences in the ranges of $P_c$ values stabilized by being out of chemical equilibrium means that the different cases can still be distinguished.

\subsection{Two-phase nonbarotropic EoS}

When computing the fundamental radial modes of the stars with the two-phase EoS, we solved Eq.~(\ref{eq:dxidr2}--\ref{eq:dDeltaPdr2}) while imposing the boundary conditions Eq.~(\ref{eq:OuterBC}) and Eq.~(\ref{eq:CenterBC}) with $\Gamma\rightarrow\Gamma_1$. At the phase transition we imposed the slow, rapid and reactive junction condition, with the latter given by Eq.~(\ref{eq:GenJunction3}) and~(\ref{eq:ReactiveJC}). In using Eq.~(\ref{eq:ReactiveJC}) we took $Y\rightarrow Y_s$ as it is the species fraction which is zero below the transition and nonzero above it. We considered the crust part of the EoS as barotropic, setting $\Gamma_1=\Gamma$ and ignoring species fraction gradients there.

The reactive junction condition required a value for parameter $Z$. Since the physics of the deconfinement transition is not fully understood, instead of computing $Z$ microphysically we choose different values for this parameter within the range $-(\rd Y_s/\rd\ln n_b)_+\leq Z\leq 0$. A typical value for $(\rd Y_s/\rd\ln n_b)_+$ for our equation of state is $\approx -0.0097$.

\begin{figure}
\includegraphics[width=\columnwidth]{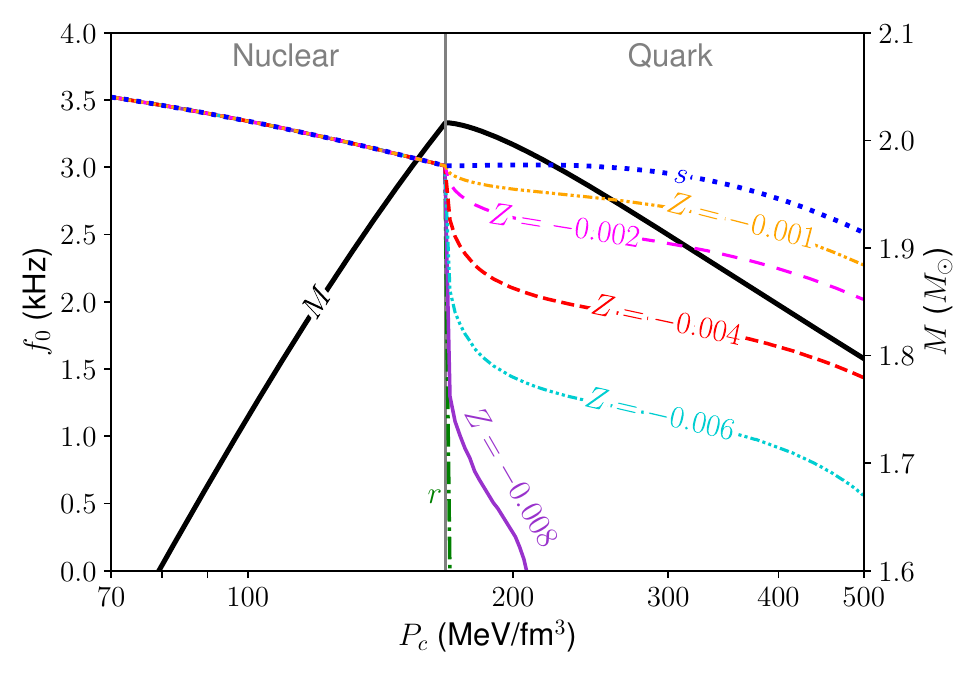}
\caption{Fundamental radial mode frequency $f_0$ (left) and stellar mass $M$ (right) as functions of central pressure $P_c$ for the nuclear plus quark matter EoS star. $f_0<0$ corresponds to an imaginary (unstable) frequency. Mode frequencies are labeled by the junction conditions used: $s$ for slow, $r$ for rapid, and $Z$ for the reactive junction condition with the value of the $Z$ parameter used shown. The different values of $-(\rd Y_s/\rd\ln n_b)_+\leq Z\leq 0$ interpolate between the rapid and slow cases. A solid vertical line indicates the central pressure corresponding to the phase transition (gray), which occurs at almost the same central pressure as that for the maximum mass star. The phase in the center of the star is labeled at the top of the plot.}
\label{fig:ReactiveJCModes}
\end{figure}

FIG.~\ref{fig:ReactiveJCModes} shows the fundamental radial mode frequency $f_0$ for the different choices of junction condition. This clearly demonstrates that the rapid and slow cases are the limiting cases of the reactive junction condition, which interpolates between these two cases depending on the value of $Z$. As was found in RS23 for the intermediate junction condition Eq.~(\ref{eq:IntermediateSpeedJunctionConditions}) for $0\leq \alpha< 1$, any change of $Z$ away from $-(\rd Y_s/\rd\ln n_b)_+$ results in a star with some previously unstable central pressure being stabilized. In this sense, the reactive junction condition is similar to the slow case, even if a smaller range of central pressures in the BTM-unstable range is stabilized compared to the slow case. 

Stars with a rapid phase transition will support a reaction mode~\citep{Haensel1989}, a radial mode that does not correspond to a mode of the single-phase star, resulting in a discontinuity in the mode frequency for fixed radial node number. The reaction mode is often, but not always, the fundamental mode. When using the reactive junction condition, the case $Z\neq 0$ also supports a reaction mode. This is clearly shown by examining FIG.~\ref{fig:ReactiveJCModes} and noting the discontinuities in $f_0$ at the phase transition when the reactive junction condition is imposed: this is most pronounced for $Z=-0.008,-0.006$. The fundamental modes are not the reaction mode for the other values of $Z$, so for these values a higher-order harmonic is the reaction mode.

\begin{figure}
\includegraphics[width=\columnwidth]{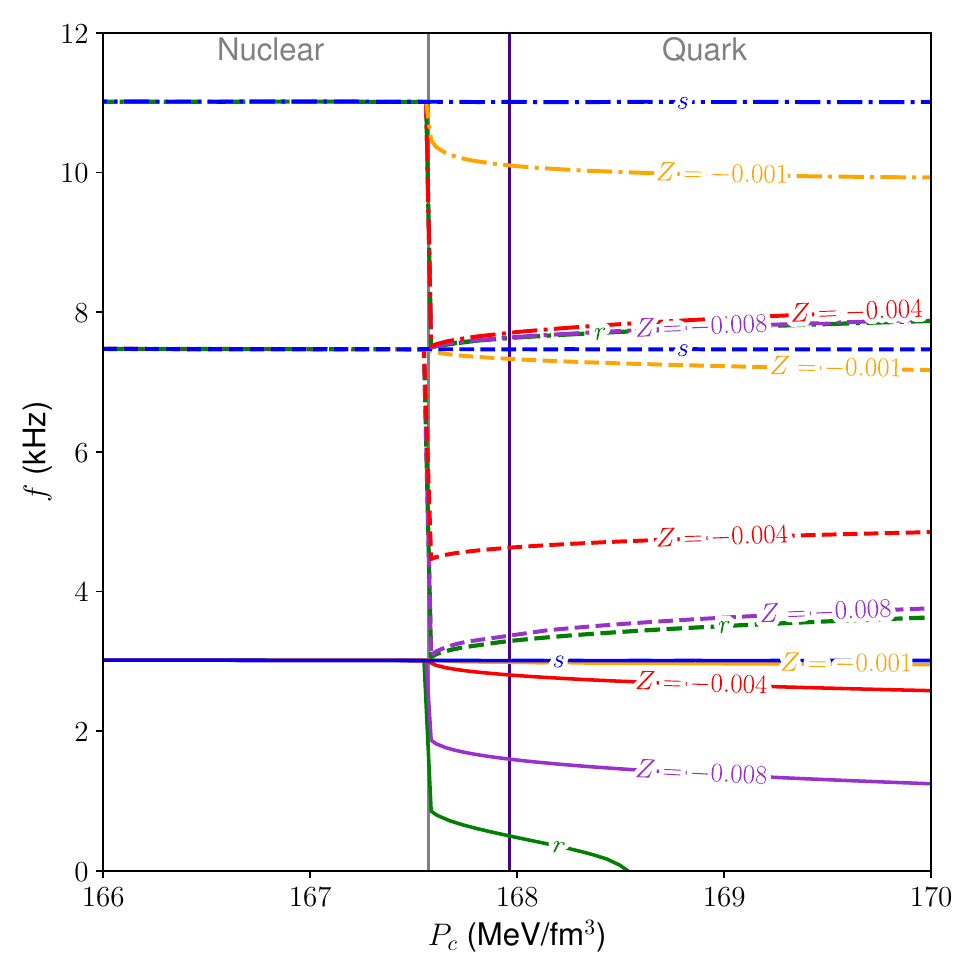}
\caption{Radial mode frequency $f$ for fundamental and first two harmonic modes as a function of central pressure $P_c$ for the nuclear plus quark matter EoS star and the reactive junction condition. Mode frequencies are labeled $r$ (rapid junction condition), $s$ (slow junction condition) or the value of $Z$ used with the reactive junction condition, with solid, dashed and dot-dashed lines for the fundamental, first harmonic and second harmonic modes respectively. Solid vertical lines are placed at the central pressures corresponding to the phase transition (gray) and the maximum mass (indigo). Note that some modes are nearly overlapping.}
\label{fig:ReactionModes}
\end{figure}

To illustrate this point, in FIG.~\ref{fig:ReactionModes} we show the fundamental and first and second harmonic radial modes for the reactive junction condition with different values of $Z$. The plot is zoomed in to central pressures zoomed in near the phase transition, and the results using the rapid and slow junction conditions are also shown for comparison. Note that $f_0$ for the rapid junction condition does not become imaginary at exactly the maximum mass because of the non-equilibrium effect $\Gamma_1\neq\Gamma$, though the range of $P_c$ with decreasing stellar mass that correspond to stable stars is so small that it is only visible in this zoomed-in plot and not in FIG.~\ref{fig:ReactiveJCModes}. The reaction mode is clearly the fundamental mode for the rapid junction condition and the reactive junction condition with $Z=-0.008$, but it is the first harmonic for the reactive junction condition with $Z=-0.004$ and the second harmonic for the reactive junction condition with $Z=-0.001$. Since the reaction mode effectively slots into the usual mode spectrum and pushes the other modes to higher frequency, this suggests that in the slow limit the reaction mode is raised to an infinitely high harmonic.

\begin{figure}
\includegraphics[width=\columnwidth]{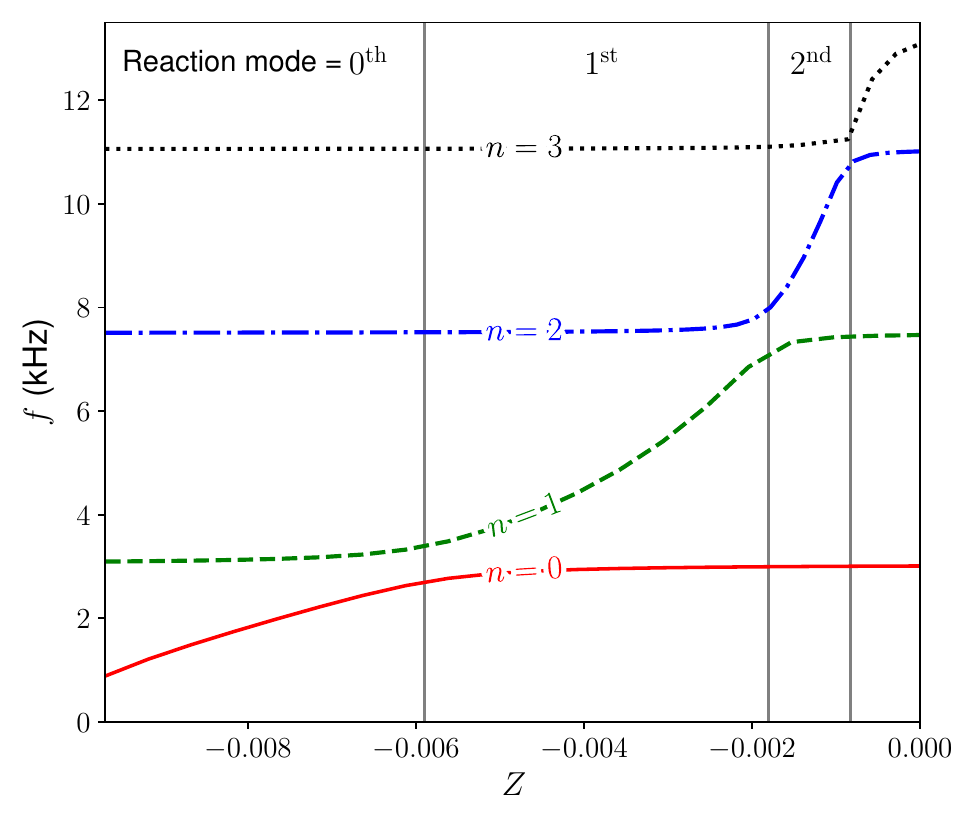}
\caption{Radial mode frequency $f$ for fundamental and first three harmonic modes as a function of the parameter $Z$ in the reactive junction condition at $P_c=167.604$ MeV/fm$^3$. Curves are labeled by the number of radial nodes: $n=0$ is the fundamental mode, $n=1$ the first harmonic, etc. $Z=0$ is the slow junction condition and $Z\approx -0.0097$ is the rapid junction condition. The avoided crossings between the curves define the ranges of $Z$ for which the different modes are the reaction mode; the harmonic corresponding to the reaction mode in each range of $Z$ defined by vertical grey lines is indicated at the top ($0^{th}=$ fundamental).}
\label{fig:ReactionModesZvsf}
\end{figure}

To clarify this point further, in FIG.~\ref{fig:ReactionModesZvsf} we plot the frequencies of the fundamental and first three harmonic modes as a function of $Z$ for fixed $P_c$ just above the phase transition. The avoided crossings between the frequency curves for constant radial node number $n$ define the ranges of $Z$ for which the reaction mode is a particular mode. For $Z$ less than $\approx-0.006$, the value corresponding to the avoided crossing between the $n=0$ and $n=1$ curves, the fundamental mode is the reaction mode; for $Z$ between $\approx-0.0059$ and $\approx-0.0018$, the avoided crossing between the $n=1$ and $n=2$ curves, the first harmonic mode is the reaction mode, etcetera. 

\section{Conclusion}
\label{sec:Conclusion}

We have studied the effects of non-equilibrium physics on the stability of hybrid stars with first-order phase transitions. In the first example, we used a three-phase barotropic EoS with the two denser quark phases modeled using constant equilibrium sound speeds. Here the out-of-equilibrium physics was modeled by choosing values for the adiabatic sound speed greater than the equilibrium sound speeds in the quark phases. This results in stable hybrid stars over an extended range of central pressures compared to the always-equilibrated case, a result consistent with studies of out-of-chemical equilibrium stability of white dwarfs and neutron stars. This extended range of stability permits the existence of higher-order stellar multiplets than those that are BTM-stable even in the case of only rapid phase transitions, since stars with central pressures below/above those corresponding to the local minima/maxima of the stellar mass $M$ are stabilized. For the EoS we examined, this permitted stable quadruplet stars with rapid phase transitions when the BTM criterion would have predicted only stable triplet stars.

In the second part of this paper, we have introduced a new junction condition to be applied when computing the radial normal modes of a compact hybrid star with strong first-order phase transitions. We have shown that this reactive junction condition interpolates between the slow and rapid junction conditions as limiting cases, but also satisfies the generalized form of junction conditions for radial oscillations of a relativistic star. It can only be applied with an equation of state that does not assume chemical equilibration and includes explicit chemical fraction-dependence. We chose a two-phase EoS with nuclear and deconfined quark matter separated by a first-order phase transition to apply this new junction condition. For different values of parameter $Z$, which is a function of the rate of particle creation (for our model, the strange quarks), we showed that it interpolates between the slow and rapid limiting cases, providing a more physically reasonable junction condition. We also showed that the reaction mode which appears in the radial mode spectrum when using the rapid junction condition persists when using the reactive junction condition, and becomes a higher harmonic as the parameter $Z$ is made smaller in magnitude (less negative).

Extensions of this work include the application of the reactive junction condition to a hybrid star with multiple phase transitions-- it could not be applied to the three-phase EoS we used in the first part of the paper because that EoS assumed chemical equilibrium. The parameter $Z$ appearing in the reactive junction condition depends on the physics of the deconfinement transition which we did not examine in detail. Computing it microscopically and using this value would allow a realistic determination of the stabilized range of $P_c$ for a given EoS. The physics of deconfinement could thus be constrained via the observation of reactively-stabilized compact stars with lower masses and radii than those observed for the maximum mass star.

\section*{Acknowledgements}

This work was supported by the Institute for Nuclear Theory's U.S. Department of Energy grant No. DE-FG02-00ER41132. G.~G.~S. participated in the National Science Foundation-funded LSAMP program at the University of Washington (Grant No. EES-1911026) while working on this paper. P.~B.~R. would like to thank S. Reddy, S.~P. Harris and T. Zhao for helpful discussion. We also thank the anonymous referee for helpful comments. All plots were made using the Python package \texttt{matplotlib}~\citep{Matplotlib}.

\bibliographystyle{apsrev4-2}
\bibliography{library,librarySpecial,textbooks}

\end{document}